\documentclass[journal]{IEEEtran}
\ifCLASSINFOpdf
\else
\fi
\usepackage{bm}  
\usepackage{multicol}
\usepackage{multirow}
\usepackage{placeins}
\usepackage{graphicx}
\usepackage{amsmath}
\usepackage{array}
\newcolumntype{H}{>{\setbox0=\hbox\bgroup}c<{\egroup}@{}}
\usepackage{graphicx}
\usepackage{caption}
\usepackage{subcaption}
\usepackage{float}
\usepackage{tabularx}
\usepackage{arydshln}
\usepackage[autostyle]{csquotes}
\usepackage{color}

\usepackage[utf8]{inputenc}
\usepackage[english]{babel}

\usepackage{amsthm}

\usepackage{hyperref}
\usepackage{algorithm}
\usepackage{algpseudocode}

\usepackage{chngpage}
\usepackage[compress]{cite}

\def\blackslug{\hbox{\kern1pt\vrule height6pt width4pt  depth1pt\kern1pt}}

\def\edp{\penalty 500\hbox{\quad\blackslug}\ifmmode\else\par
	\vskip4.5pt plus3pt minus2pt\fi}

\newtheorem{reduction test}{Reduction Test}

\begin{document}
%

%
\title{Physarum-inspired Network Optimization: A Review}

\author{\IEEEauthorblockN{Yahui Sun} \\
	\IEEEauthorblockA{\url{https://yahuisun.com}}
}
%
	\maketitle
	\begin{abstract}
		The popular Physarum-inspired Algorithms (PAs) have the potential to solve challenging network optimization problems. However, the existing researches on PAs are still immature and far from being fully recognized. A major reason is that these researches have not been well organized so far. In this paper, we aim to address this issue. First, we introduce Physarum and its intelligence from the biological perspective. Then, we summarize and group four types of Physarum-inspired networking models. After that, we analyze the network optimization problems and applications that have been challenged by PAs based on these models. Ultimately, we discuss the existing researches on PAs and identify two fundamental questions: 1) What are the characteristics of Physarum networks? 2) Why can Physarum solve some network optimization problems? Answering these two questions is essential to the future development of Physarum-inspired network optimization. 
	\end{abstract}

	
	%
	\section{Introduction}
	
	The emerging next-generation network optimization techniques are expected to solve challenging problems in various areas, such as transportation, communication, and biomedical data analytics. After millions of years of evolution, the bio-intelligence may provide clues to deign such techniques. In fact, many types of bio-intelligence have been exploited to solve various network optimization problems, and they have considerably improved the strategies currently available in our research community. However, some network optimization problems still remain unsolved, and new techniques are required to address this issue.
	
	Physarum polycephalum is a slime mold that inhabits shady, cool and moist areas (see Figure \ref{ppp1}; for the convenience, we simply refer to Physarum polycephalum as Physarum in this paper). In previous biological experiments, it has exhibited an extraordinary intelligence to build efficient networks, such as the shortest paths \cite{imba2000} and the Steiner trees \cite{amib2010, copf2010, omsf2004}. Hence, an increasing number of researches have recently been conducted to explore this intelligence for network optimization. 
	
	A lot of Physarum-inspired Algorithms (PAs) have been proposed to challenge network optimization problems, such as the shortest path problem \cite{psab2006,fspo2012}, the traveling salesman problem \cite{auos2014}, and the Steiner tree problems \cite{faib2016,apps2016}. The ability of PAs to solve some network optimization problems has already been proven theoretically. For example, Bonifaci et al. \cite{pccs2012} proved that the PA proposed by Tero et al. \cite{ammf2007} can compute the shortest paths independently of the network topology. Nevertheless, most of the existing PAs are still lack of solid theoretical bases, even though they have shown a more competitive performance than some traditional techniques. Furthermore, the existing researches on PAs have not been well organized to date. Therefore, the development of Physarum-inspired network optimization is still immature and far from being fully recognized, and some fundamental questions still remain unanswered today. In this paper, we will address this issue by reviewing and discussing the exiting researches on Physarum-inspired network optimization.
	
	The structure of this paper is as follows: in Section \ref{Section: Physarum and its intelligence}, we introduce Physarum and its intelligence; in Section \ref{Section: The proposed models of Physarum-inspired network optimization}, we summarize and group four types of Physarum-inspired networking models; in Section \ref{Section: The challenged network optimization problems and applications}, we summarize the network optimization problems and applications that have been challenged by PAs based on these models; in Section \ref{Section: Discussion}, we discuss the exiting researches on PAs and then provide some recommendations for future work.

	\section{Physarum and its intelligence} \label{Section: Physarum and its intelligence}
	
	In this section, we first introduce Physarum from the biological perspective. Then, we summarize its intelligent behaviors. Ultimately, we reveal the physiological mechanism in Physarum that possibly accounts for this intelligence.
	
	\subsection{Physarum polycephalum}
	
\begin{figure}	[!t]
		\centering
		\begin{subfigure}[t]{0.475\textwidth}
			\centering
			\includegraphics[width=\textwidth,height=0.12\textheight]{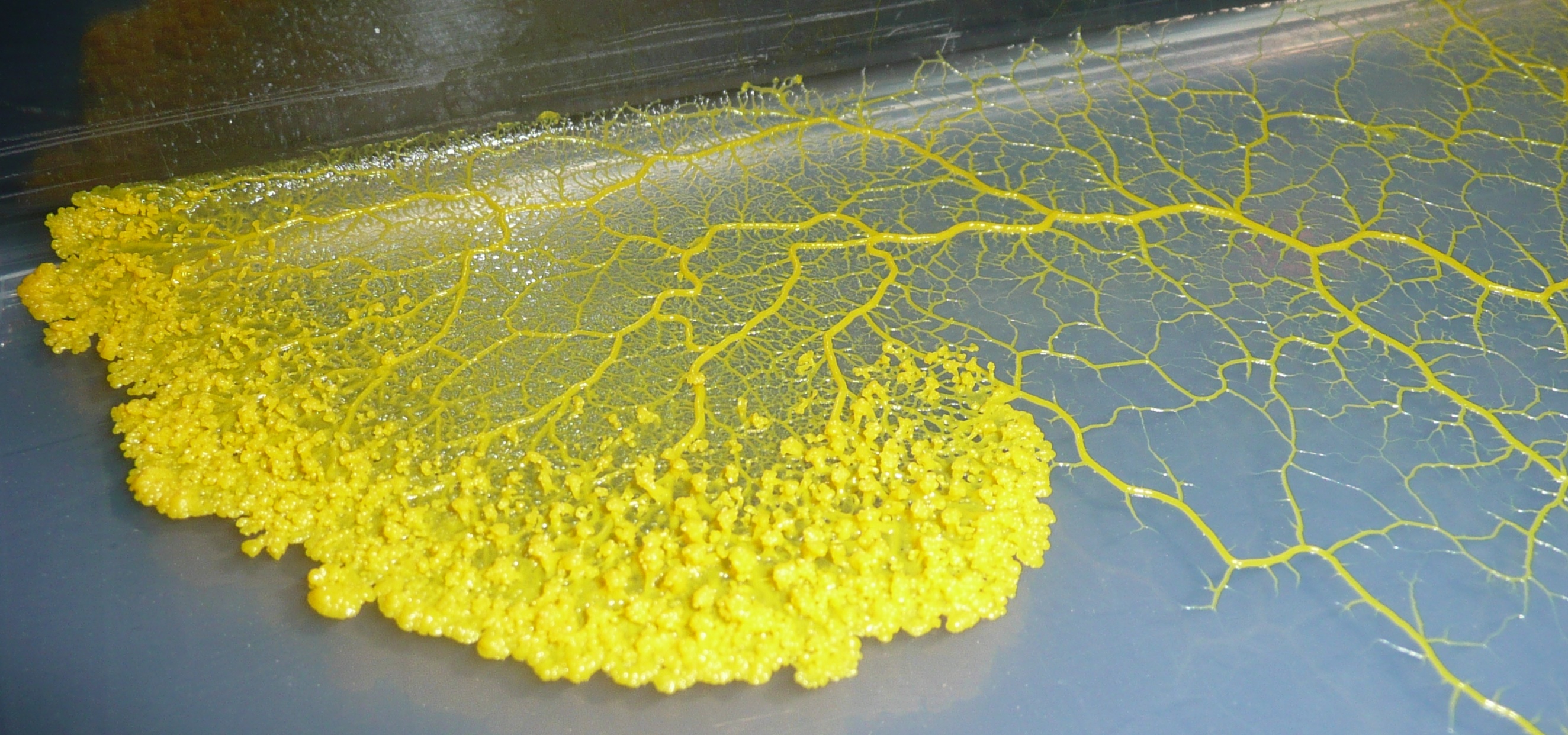}
			\subcaption{The tubular structure of Physarum polycephalum}
			\label{ppp1}	
		\end{subfigure}
		\hfill
		\begin{subfigure}[t]{0.475\textwidth}
			\centering
			\includegraphics[width=\textwidth]{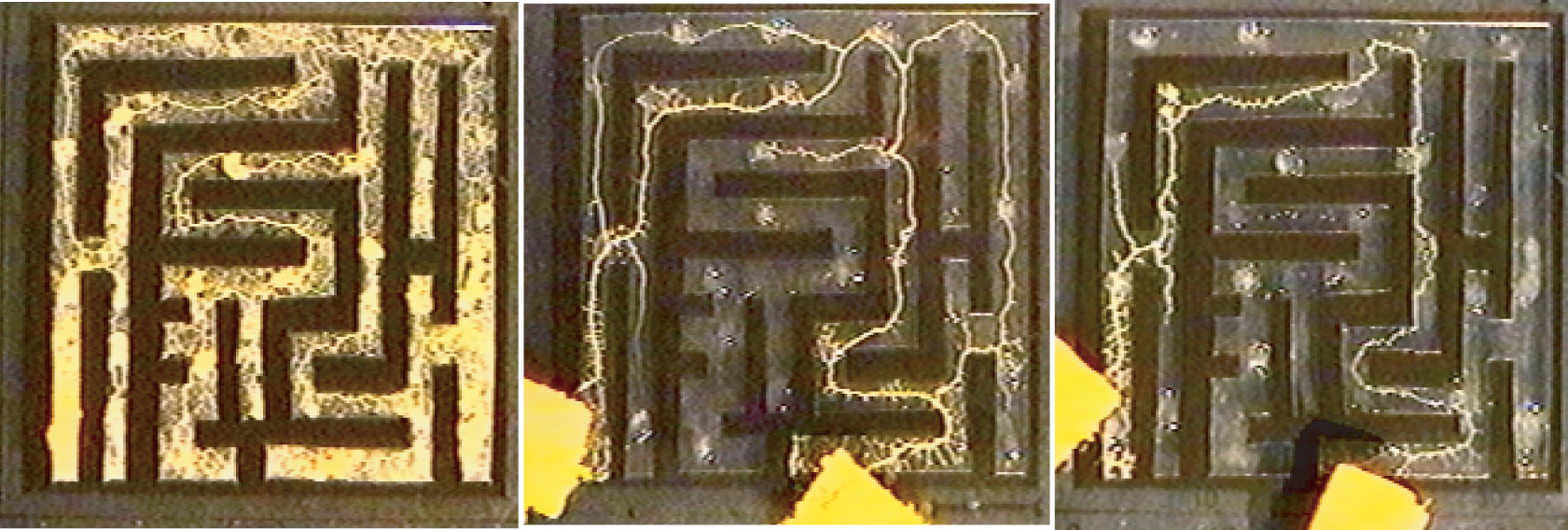}
			\subcaption{Physarum polycephalum finds the shortest path}
			\label{ppp2}
		\end{subfigure}
		\hfill
		\begin{subfigure}[t]{0.475\textwidth}
			\centering
			\includegraphics[width=\textwidth]{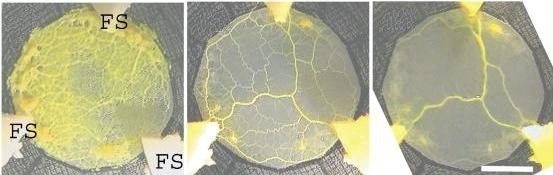}
			\subcaption{Physarum polycephalum forms a Steiner tree}
			\label{ppp3}
		\end{subfigure}
		\caption{Photographs of Physarum polycephalum (provided by Professor Toshiyuki Nakagaki from Hokkaido University). (a) shows the tubular structure of Physarum polycephalum. (b) shows an experiment in which Physarum polycephalum finds the shortest path between two agar blocks in a maze. (c) shows an experiment in which Physarum polycephalum forms a Steiner tree to connect three food sources (FS).}
		\label{ppp photo}
	\end{figure}

	Physarum polycephalum is a species of order Physarales, subclass Myxogas-tromycetidae, class Myxomecetes, division Myxostelida, commonly known as a true slime mold \cite{dpgb2009}. This organism has a sophisticated life cycle, which was first described by Howard in 1931 \cite{tlho1931}. At some stage of the cycle, it becomes a syncytial mass of protoplasm, called plasmodium, which is a single cell with many
	diploid nuclei that has a size of up to tens of centimeters. The plasmodium of Physarum polycephalum consists of two parts (see Figure \ref{ppp1}): a \enquote{sponge} section including distributed actin-myosin fibers, and a \enquote{tube} section made up of actin-myosin fibers. These two parts consist of a mycelial network, which acts both as an information highway transporting chemical and physical signals and as a supply network circulating nutrients and metabolites throughout the organism. The topology of this network changes as Physarum explores the neighboring environment. Many remarkable intelligent behaviors have been displayed during this process, and it is believed that this intelligence can be exploited to solve various network optimization problems.

	\subsection{Physarum intelligence}
	
	Physarum has shown many intelligent behaviors. We summarize these behaviors as follows:
	
	\subsubsection{Finding the shortest path}
	
	this intelligent behavior was first observed by Nakagaki et al. in 2000 \cite{imba2000}. In their experiment, Physarum successfully found the shortest path between two selected points in a maze (see Figure \ref{ppp2}). Notably, people sometimes refer to this ability as the maze-solving ability \cite{sedw2013}. Moreover, it may be worthy mentioning that Physarum can solve many other well-known problems that can be transformed to the shortest path problem, such as the minimum-risk problem \cite{mpfb2007} and the towers of Hanoi problem \cite{bbmd2015}.
	
	\subsubsection{Building high-quality networks}
	
	Physarum can build high-quality networks to connect multiple food points. After millions of years of Darwinian natural selection, it is believed that Physarum networks  have achieved a good balance between cost, efficiency and resilience. For example, in a famous experiment operated by Tero et al. in 2010 \cite{rfbi2010}, Physarum built networks with comparable qualities to those of Tokyo rail system. Some other real-world transportation networks have also been approximated by Physarum since then, such as the Mexican and Iberian highways \cite{amhw2011,rimw2011}. More remarkably, many other experiments have shown that the tubular topologies of Physarum are sometimes similar to those of complex mathematical networks (eg. the Steiner trees in Figure \ref{ppp3}) \cite{eomo2007,abn2009,maao2014,osco2009,rosm2015}. This type of Physarum intelligence is highly valued, especially when considering the difficulty of using traditional techniques to design such networks.

	\subsubsection{Adapting to changing environments}
	
	Physarum is chemotaxis, phototaxis and thermotaxis. Previous biological experiments have shown that Physarum networks disassemble and reassemble within a period of a few hours in response to the change of external conditions \cite{ammf2007,rmis2010}. For instance, when chemicals are applied to any part of the body of Physarum, the whole organism migrates towards or away from the stimulus \cite{amop1998}. More radically, Adamatzky \cite{spwl2009} have shown that Plasmodium-based computing devices can be precisely controlled and shaped by illumination. This adapting ability is highly appreciated in the design of dynamic networks, such as the mobile ad hoc networks \cite{bnft2010,and2010}.
	
	\subsubsection{Memorizing and learning}
	
	as a unicellular organism, Physarum has shown an amazing ability to memorize and learn. This intelligent behavior was first revealed by Saigusa et al. in 2008 \cite{aape2008}. In their experiment, Physarum was initially exposed to unfavorable conditions presented as three consecutive pulses at constant intervals. In response, Physarum reduced its locomotive speed in each pulse. Then, the unfavorable conditions were removed. However, Physarum still reduced its locomotive speed at the time when the unfavorable pulse would have occurred. Therefore, this experiment shows that Physarum can memorize and learn. Moreover, Shirakawa et al. \cite{aale2011} used an associate learning experiment to further test this ability, and Reid et al. \cite{smua2012} pointed out that Physarum can leave a thick mat of non-living, translucent, extracellular slime as a form of spacial memory to avoid areas where it has explored.
	
	\subsubsection{Biological computing}
	
	Physarum is considered to be amongst the most prospective experimental prototypes of biological computers. Many experiments have been done to exploit its computing ability. For example, Tsuda et al. \cite{raep2004} used Physarum to make a Boolean gate; they further used Physarum to control robots in unknown dynamic environments \cite{rcwb2007}; Adamatzky et al. \cite{prop2010} made a programmable Physarum machine.
	
	\subsubsection{Distributed intelligence}
	
	in the body of Physarum, there is no central information processing unit like a brain, but rather a collection of similar parts of protoplasm. As a consequence, a piece of Physarum cut from a larger one can regenerate and become a perfect organism \cite{sbot2001}, and two independent organisms can combine to form a single organism after they contact with each other \cite{maao2014}. Thus, Physarum is a good material for the researches on autonomous distributed network optimization \cite{amop1998}. As the scales of the next-generation networks are expected to be extremely large, centralized control of communication becomes impractical. With the distributed intelligence, Physarum may inform the design of next-generation, adaptive, robust spatial infrastructure networks with decentralized control systems \cite{bstt2007,apat2013}. 
	
	\subsection{The source of Physarum intelligence}
	
	It is necessary to understand where the Physarum intelligence comes from before exploiting it. Since we focus on Physarum-inspired network optimization in this paper, we are interested in understanding what accounts for Physarum's ability to design high-quality networks. There is no exact answer to this question so far. Nevertheless, the evidence to date suggests that the protoplasmic flow through Physarum's tubular veins plays an important role in developing its networks. This protoplasmic flow is sometimes known as shuttle streaming since the direction of the flow changes back and forth periodically, based on a gradient of hydrostatic pressure produced by active rhythmic contraction throughout the cell \cite{acmw2005,rnpi2013}. The protoplasmic flow brings important nutrients absorbed from food sources into the tubular vein network. The more concentration of nutrients in a tube, the more likely it can absorb nutrients into its walls, and hence the more likely that it grows to a wider tube. This growth leads to a further increase of flux because the resistance to the flow is smaller in wider tubes. On the other hand, shorter tubes needs fewer nutrients to grow. Therefore, thick short tubes are the most effective tubes for nutrition transportation in Physarum. The preference of such tubes by the protoplasmic flow is believed to account for Physarum's intelligence to design high-quality networks. We will later show that the simulation and reproduction of this protoplasmic flow is the basis of many successful Physarum-inspired networking models.
	
	\begin{figure}	[!t]
		\centering
		\includegraphics[width=0.45\textwidth]{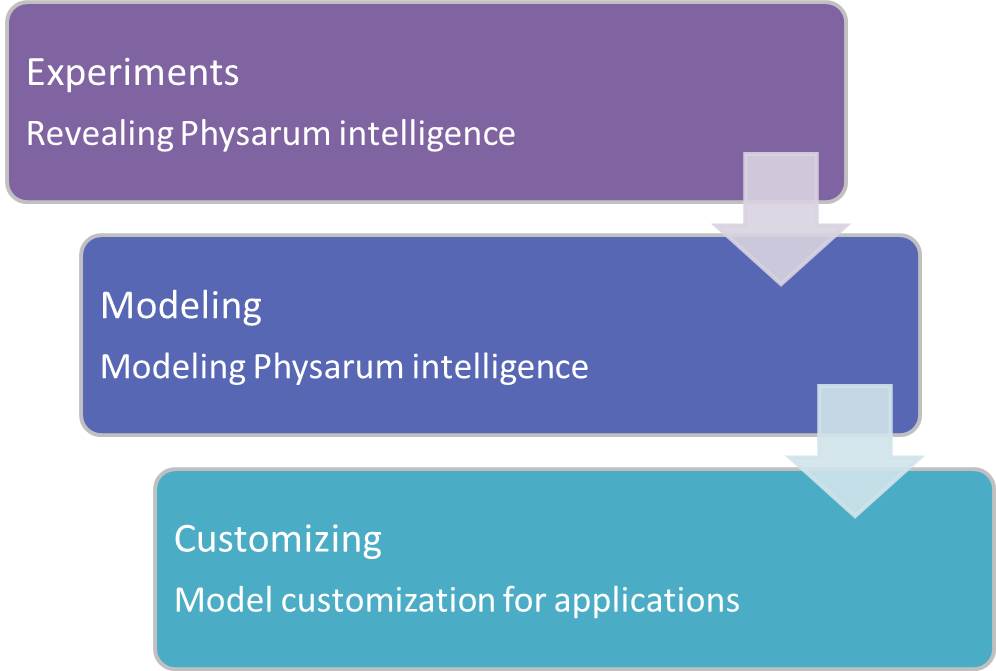}
		\label{3stps}
		\caption{Three steps to develop Physarum-inspired networking models.}
		\label{3stps photo}
	\end{figure}

	\section{The Physarum-inspired networking models} \label{Section: The proposed models of Physarum-inspired network optimization}
	
	In this section, we first introduce the methodology to develop Physarum-inspired networking models. Then, we summarize four types of Physarum-inspired networking models. We will later discuss the network optimization problems and applications that have been challenged by these models.
	
	\subsection{The methodology of developing Physarum-inspired networking models}
	
	Various bio-inspired techniques have been developed in the last few decades, including the genetic algorithm \cite{cmig1970}, the particle swarm optimization algorithm \cite{pso1995}, and the currently popular neural networks \cite{alco1943}. The methodology of developing Physarum-inspired networking models is similar to that of other bio-inspired techniques. Dressler and Akan \cite{bnft2010} have summarized three steps to develop bio-inspired networking techniques, which are identification of analogies in nature, modeling of realistic biological behaviors, and model customization for applications. Based on their work, we summarize three steps to develop Physarum-inspired networking models in Figure \ref{3stps photo}. First, it is necessary to reveal the Physarum intelligence of network optimization through biological experiments. Second, we can model Physarum intelligence based on the observation in these experiments. Ultimately, the proposed models must be customized for different network optimization applications.
	
	\subsection{The flow-conductivity model}
	
	The protoplasmic flow through Physarum's tubular veins is believed to account for Physarum's intelligence. Many models have been proposed to simulate this flow, in which the flow-conductivity model is the most successful one. This model was first proposed by Tero et al. in 2006 \cite{psab2006} to solve the shortest path problem, and it is inspired by an underlying physiological mechanism: Physarum's tube thickens as the protoplasmic flow through it increases \cite{sbot2001}.
	
	The shortest path problem is to find the shortest path between two terminals in a network. In this model, there is protoplasmic flow in every edge. The two terminals represent two agar blocks containing nutrient, which are food for Physarum. One terminal is called the source node, and the other terminal is called the sink node. The protoplasmic flow flows into the network from the source node and out of the network from the sink node. There is pressure at each vertex, and the quantity of flux in each edge is proportional to the pressure difference between the two ends of this edge. Specifically, the flux $Q_{ij}$ in edge $(i,j)$ is given by the Hagen-Poiseuille equation below.
	\begin{eqnarray} \label{calQ}
		Q_{ij} = \frac{D_{ij}}{c_{ij}} (p_i - p_j)
	\end{eqnarray}
	\begin{eqnarray} \label{D_r}
		D_{ij} = \frac{\pi r^4_{ij}}{8 \xi}
	\end{eqnarray}
	\noindent where $D_{ij}$ is the edge conductivity, $c_{ij}$ is the edge length, $p_i$ and $p_j$ are pressures at vertices $i$ and $j$, $r_{ij}$ is the edge radius, and $\xi$ is the viscosity coefficient. Equation (\ref{D_r}) shows that Physarum's tubular thickness ($r_{ij}$) increases with the tube's conductivity. Thus, the change of Physarum's tubular thickness can be described by the conductivity update equation as follows.
	\begin{eqnarray} \label{updateD}
		\frac{d}{dt} D_{ij} = f(|Q_{ij}|) - \mu D_{ij}
	\end{eqnarray}
	where $f(|Q_{ij}|)$ is an increasing function, $\mu$ is a positive constant. The conductivity update equation implies that conductivities tend to increase in edges with big flux. Therefore, the conductivity update equation reflects the physiological mechanism above. To calculate the flux and update edge conductivities, we need to first calculate the pressures. By considering the conservation law of flux at each vertex, the pressures can be calculated using the network Poisson equation below.
	\begin{eqnarray} \label{old pressure}
		\sum_{i \in V(j)} \frac{D_{ij}}{c_{ij}} (p_i - p_j) =\left\{
		\begin{array}{c l}      
			-I_0, & j=source\\
			+I_0, & j=sink\\
			0, & otherwise
		\end{array}\right.
	\end{eqnarray}
	where $V(j)$ is the set of vertices linked to vertex $j$, and $I_0$ is the quantity of flux flowing into the source node and out of the sink node. Let the pressure at the sink node be 0, and give each edge conductivity an initial value, then the other pressures can be calculated using Equation (\ref{old pressure}). After that, the quantity of flux in each edge can be calculated using Equation (\ref{calQ}), and the conductivity of each edge can be updated using Equation (\ref{updateD}). Given a threshold value of edge conductivity, edges with conductivities smaller than this value are cut from the network. In this way, the shortest path between the source node and the sink node can be found by iteratively updating edge conductivities and cutting edges.
	
	Miyaji et al. \cite{pcst2008} provided a rigorous proof that the equilibrium point corresponding to the shortest path is globally asymptotically stable for the model above on Riemannian surface. Later, Bonifaci et al. \cite{pccs2012} proved that this model converts to the shortest path, independently of the network topology. However, it must be noticed that both of their proofs are based on the preconditions that $f(|Q_{ij}|)=|Q_{ij}|$ and $\mu =1$.
	
	The initial flow-conductivity model above was designed to find the shortest path between two terminals. Nevertheless, there are many adapted models to deal with the multi-terminal cases, such as the Steiner tree problems. The biggest challenge of adapting the flow-conductivity model to the multi-terminal cases is how to select terminals to be source and sink nodes. There are four strategies to do so:
	
	\begin{itemize}
		\item  \textbf{One source node and one sink node:} this strategy is to select one terminal to be the source node and then select another terminal to be the sink node. It was first proposed by Nakagaki et al. in 2008 \cite{caoc2008} to solve the classical Steiner tree problem in planes. Some later researches are also based on this strategy. For example, Tero et al. \cite{rfbi2010} designed networks similar to the Tokyo rail system, Houbraken et al. \cite{ftnd2013} designed fault tolerant networks, and Qian et al. \cite{aacs2013} solved the traveling salesman problem.
		
		\item \textbf{Multiple source nodes and one sink node:} this strategy is to select one terminal to be the sink node and then select the other terminals to be source nodes. It has been applied by Liu et al. in 2015 \cite{poab2015} to solve the classical Steiner tree problem in graphs. Recently, Sun et al. \cite{faib2016,apps2016} have also used it to solve the prize-collecting Steiner tree problem and the node-weighted Steiner tree problem.
		
		\item \textbf{One source node and multiple sink nodes:} this strategy is to select one terminal to be the source node and then select the other terminals to be sink nodes. It was first used by Watanabe et al. in 2014 \cite{tnwf2014} to design transportation networks with fluctuating traffic distributions. Later, Liu et al. \cite{abii2016} also used it to identify the focal nodes that spread diseases in epidemiological networks.
		
		\item \textbf{Multiple source nodes and multiple sink nodes:} this strategy is to select multiple terminals to be the source nodes and multiple terminals to be the sink nodes. It was recently proposed by Zhang et al. in 2016 \cite{aips2016} to solve the supply chain network design problem. 
	\end{itemize}

	On the other hand, some work has been done to modify the flow-conductivity model. For instance, Zhang et al. accelerated its optimization process by intentionally removing the edges with a stable decreasing flow \cite{rpaf2014}; Tero et al. \cite{ammf2007} analyzed the influence of different forms of $f(|Q_{ij}|)$ through abundant computational experiments; Liu et al. \cite{poab2012} proposed an equation to obtain new pressures by updating old pressures, and thus avoided the computationally expensive process of calculating pressures. Moreover, there are some other models that are similar to the flow-conductivity model, such as the current-reinforced random walk model \cite{crwf2013}. 
	
	\subsection{The cellular model}
	
	The cellular model was proposed by Gunji et al. in 2008 \cite{mmoa2008} to design high-quality networks. In this model, the optimization process is consistent with the properties of real cells. There are two phases in this process: the development phase and the foraging phase. In the development phase, an aggregation of cell components is derived from an initial seed. While in the foraging phase, the behavior of the cell corresponds to that of the vegetative state of Physarum. Given a planer lattice, and every lattice site has various states, a cell is then described as an aggregation of lattice sites in particular states: the inside (state 1) is surrounded by a boundary (state 2) in a lattice space consisting of the outside (state 0). It is assumed that the boundary state corresponds to an assembly of cytoskeleton fibers in the body of Physarum. In the foraging phase, a cell eats 0, and this gives rise to both migration and modification of the cell. Eating 0 or invasion of the outside into a cell corresponds to the process of softening a particular part of the membrane of Physarum. The protoplasmic flow toward the softened area is implemented by the transportation of the eaten 0, which is called the bubble. During this transportation, the bubble is accompanied by cytoskeleton, which leads to a re-organization of the distribution of the cytoskeleton. 
	
	Gunji et al. \cite{mmoa2008,aaar2011} have applied this model to simulate the amoebic motion and solve the classical Steiner tree problem in planes. However, Liu et al. \cite{apne2013} pointed out that the initial cellular model is of low efficiency since there is only one bubble in the cell. Therefore, they proposed an improved cellular model with multiple bubbles, and their improved model obtained higher efficiency and stability than the original one. Moreover, Adamatzky proposed a different cellular model in 2014 \cite{r2a72014}. His model is different from the cellular model above in that it imitates the active growing zones of Physarum with excitation wavefronts and structure of protoplasmic tubes with pointers in excitable cellular automata. He has successfully applied this model to approximate the longest roads in USA and Germany on 3-D terrains.

	\subsection{The multi-agent model}
	
	Physarum has displayed the behaviors to design high-quality networks in previous experiments. Therefore, it is desirable to design a model to approximate the behaviors of Physarum. The multi-agent model proposed by Jones in 2009 \cite{atbo2009} is such a model. This model is based on the behaviors of a population of particle-like mobile agents to approximate some of the complex phenomena observed in Physarum. Each agent senses and deposits stimulus as it moves towards the nearby stimulus within a diffusive lattice, which is represented by a discrete two-dimensional floating point array. Therefore, the swarm population not only adapts to but also modifies its environment. In this model, the structure of the Physarum network is indicated by the collective
	pattern of the positions of agents, and the protoplasmic flow is represented by the collective movement of agents. Just like the biological experiments on Physarum \cite{dpgb2009}, Jones have applied the initial multi-agent model to approximate different types of proximity graphs \cite{atbo2009,copf2010,msmc2016} and construct complex transport networks \cite{tpsm2011}. Furthermore, Wu et al. \cite{aems2012} improved the initial multi-agent model by reducing the number of sensors of each agent and adding a memory module to each agent. Their improved model is more flexible and adaptive, and it approximates the behaviors of Physarum more closely.
	
	\subsection{The shuttle streaming model}
	
	The shuttle streaming model proposed by Siriwardana et al. in 2012 \cite{fspo2012} is based on the simulation of the protoplasmic flow through Physarum's tubular veins. This model simulates the bidirectional protoplasmic flow to solve the shortest path problem. In this model, the protoplasmic flow is divided into two parts: the forward flow and the backward flow. These two types of flow occur alternately from one terminal to the other terminal. There is a uniform absorption rate for each edge, and the quantity of flux each edge absorbs is in proportion to this uniform absorption rate and the length of this edge. Thus, longer edges absorbs more flow. Each node has a memory of the quantity of flux it received through each edge in the previous flow in the opposite direction, and the node distributes the flux it received in the current flow to the adjacent edges in proportion to the quantity of flux it received in the last flow. Similar to the flow-conductivity model, edges with small flow in them are cut from the network. By iterating the bidirectional protoplasmic flow above, the shortest path between two terminals can be found. Siriwardana et al. have shown that the shuttle streaming model is much faster than the flow-conductivity model to find the shortest path. Nevertheless, there is no rigorous proof that it can always find the shortest path. Moreover, no work has been done to apply it to more complex network optimization problems. Hence, future work is recommended to release the full potential of this model.

	\section{The challenged network optimization problems and applications} \label{Section: The challenged network optimization problems and applications}
	
	The above Physarum-inspired networking models have been applied to challenge various network optimization problems and applications. We summarize them as follows.

	\subsection{The shortest path problem and applications}
	
	The shortest path problem is to find the shortest path between two terminals in a network. Many PAs have challenged this problem, in which Physarum Solver \cite{psab2006} based on the flow-conductivity model is probably the most famous one. For example, Chen et al. \cite{svwt2016} applied it to spectroscopy analysis; Zhang et al. \cite{aipp2014} improved its performance; Bonifaci et al. \cite{pccs2012} proved that it always converts to the shortest path independently of the network topology. However, no traditional algorithm for the shortest path problem has been compared with Physarum Solver and its variants to date. Thus, future work is  required to show the advantages of Physarum Solver over traditional algorithms. Besides Physarum Solver and its variants, Physarum Optimization with Shuttle Streaming \cite{fspo2012}, which is based on the shuttle streaming model, has also challenged the shortest path problem. Siriwardana et al. \cite{fspo2012} claimed that this algorithm is much faster than Physarum Solver. Nevertheless, there is no rigorous proof of its ability to find the shortest path so far, and computational trials show that it may not always convert to the shortest path \cite{fspo2012}. Hence, more theoretical work is needed to reveal its ability and then improve its performance.
	
	\subsection{The classical geometric Steiner tree problem  and applications}
	
	The classical geometric Steiner tree problem is to find the shortest network to connect multiple terminals in a given geometric space. It has been widely applied to design the shortest transportation networks in numerous areas. Nakagaki et al. \cite{faip2008} have proposed a PA based on the flow-conductivity model to challenge this problem. They first constructed a dense graph to represent the Euclidean plane. Then, they applied their PA to that graph. The subnetworks obtained in this graph are composed of meandering paths. By replacing these meandering paths with straight lines, they obtain the final solution networks. In this way, they successfully found approximations to  Steiner minimum trees with up to 16 terminals. Tero et al. \cite{amib2010} further suggested that the parameter $\mu$ in Equation (\ref{updateD}) and the rules to select source and sink nodes are essential to this algorithm. Besides the work above, Gunji et al. \cite{mmoa2008} have proposed a PA based on the cellular model to challenge this problem. In their algorithm, they use cells to cover the Euclidean plane, and the cell moves to find approximations to Steiner minimum trees. However, their algorithm can only find approximations to Steiner minimum trees with up to 4 terminals. It is easy to see that all of the algorithms above can only solve the classical geometric Steiner tree problem with limited number of terminals. Moreover, they all have a low solution precision. For example, the solution precision of Nakagaki's algorithm is limited by the density of the grid, while that of Gunji's algorithm is limited by the density of cells. Thus, it is reasonable to say the current work of using PAs to challenge the classical geometric Steiner tree problem is still immature.
	
	\subsection{The classical Steiner tree problem in graphs  and applications}
	
	The classical Steiner tree problem in graphs is to find the shortest subnetwork to connect multiple terminals in a network. Song et al. \cite{apaf2012} first proposed a PA based on the flow-conductivity model to solve the minimal exposure problem in wireless sensor networks, which can be transformed into the classical Steiner tree problem in graphs. Later, Liu et al. \cite{poab2015} improved this algorithm, and their algorithm showed a performance similar to two traditional algorithms, the 1.55 worst-case ratio algorithm and the tabu search algorithm. 
	
	\begin{figure}	[!t]
		\centering
		\includegraphics[width=0.3\textwidth]{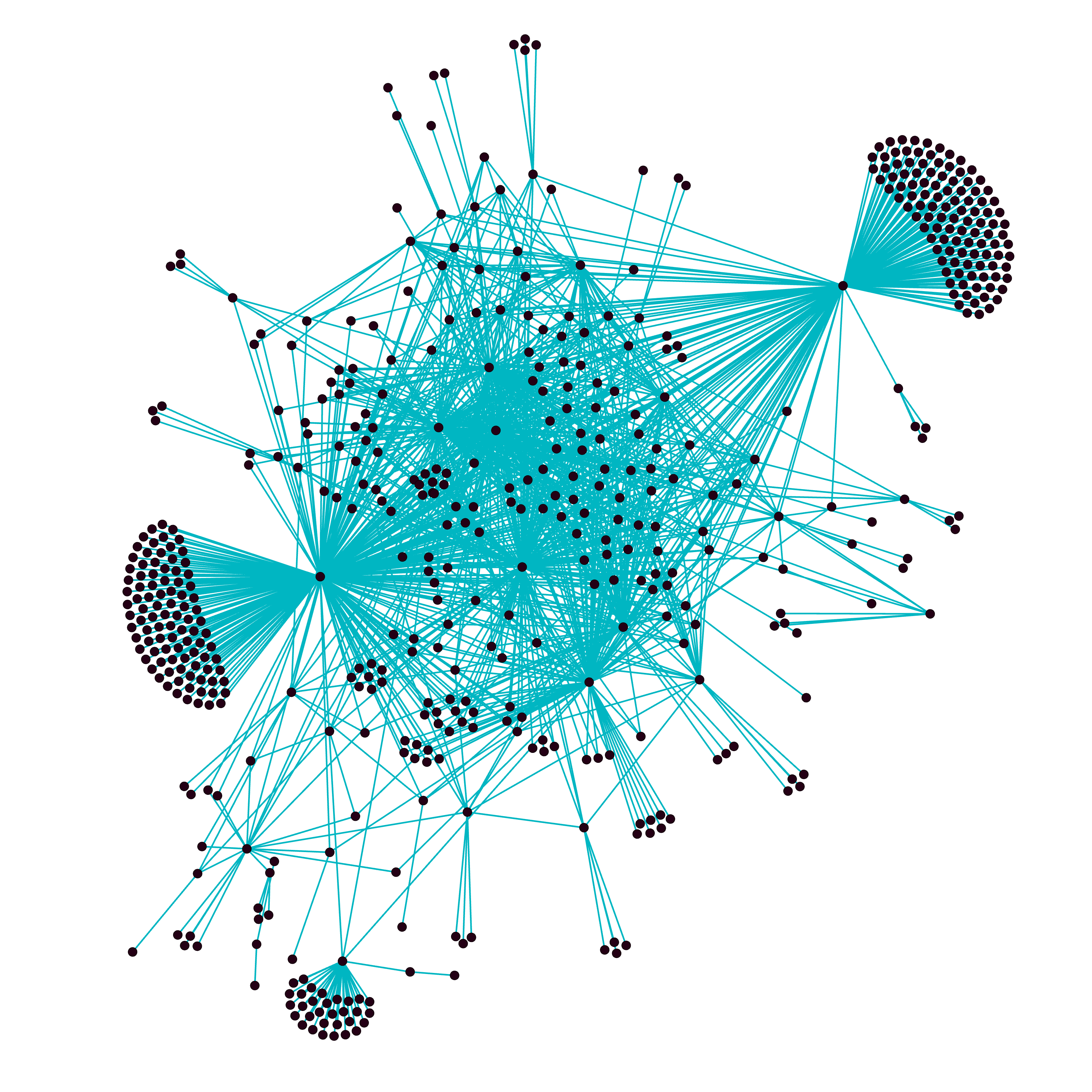}
		\caption{Application of Physarum-inspired algorithms to pharmacological networks for drug repositioning \cite{apps2016}.}
		\label{Figure: drug}
	\end{figure}
	
	\subsection{The prize-collecting Steiner tree problem  and applications}
	
	The prize-collecting Steiner tree problem is a more general version of the classical Steiner tree problem in graphs. In the prize-collecting Steiner tree problem, we not only have edge costs, but also have none-negative node weights. The purpose of this problem is to find a subnetwork with the maximum net-weight (the total node weights minus the total edge costs). Recently, Sun et al. \cite{apps2016} proposed a PA based on the flow-conductivity model to challenge this problem. They applied both their algorithm and the traditional GW algorithm to solve this problem for drug repositioning (see Figure \ref{Figure: drug}). Their algorithm successfully found more valuable drug repositioning candidates than the traditional GW algorithm. 
	
	\subsection{The node-weighted Steiner tree problem and applications}
	
	The node-weighted Steiner tree problem is a more general version of the prize-collecting Steiner tree problem, and it has been applied to various areas, such as communication network design \cite{faib2016} and biomedical data analytics \cite{tnst2017}. The difference between these two algorithms is that in the node-weighted Steiner tree problem, the node weight can be positive, negative, or zero. Recently, Sun et al. \cite{faib2016} proposed a PA based on the flow-conductivity model to challenge this problem. They compared their algorithm with two traditional algorithms, the genetic algorithm and the particle swarm optimization algorithm. In the computational trials, their algorithm found better solutions in a shorter time than these two algorithms. 
	
	\subsection{The traveling salesman problem and applications}
	
	The traveling salesman problem is to find the shortest circle connecting multiple vertices in a network. It has several applications, such as planning, logistics, and the manufacture of microchips. The evolutionary algorithms, such as the ant colony optimization algorithm, the genetic algorithm, and the particle swarm optimization algorithm, are widely used to solve this problem. These algorithms perform well in small instances. However, they may not be efficient and robust enough to solve the real-world large instances. Zhang et al. \cite{auos2014} proposed a PA based on the flow-conductivity model to update the pheromone matrix in the ant colony optimization algorithm. The final combined algorithm can avoid the condition of premature convergence which often occurs in the pure ant colony optimization algorithm. Lu et al. \cite{anpa2014} further applied this algorithm to the real-world instances with 34 vertices, and it achieved a better performance than the ant colony optimization algorithm, the genetic algorithm, and the particle swarm optimization algorithm. Besides the original traveling salesman problem, some variant problems have also been challenged by PAs. For example, Masi et al. \cite{ammp2014} and Zhang et al. \cite{maco2016} have proposed PAs to solve the multi-objective traveling salesman problem, while Pezhhan et al. \cite{abis2013} have solved the fuzzy traveling salesman problem.
	
	\subsection{The multicast routing problem and applications}
	
	The multicast routing problem is to find the lowest-cost subnetwork to deliver data to multiple destinations in a communication network. It is similar to the classical Steiner tree problem in graphs, except that some constraints are required to be met, such as the delay and bandwidth constraints. Liang et al. \cite{anga2016} proposed a PA based on the flow-conductivity model to update the crossover operator of the genetic algorithm. The final combined algorithm has challenged the multicast routing problem in mobile ad hoc networks. The computational trials show that it has a much better performance than the genetic algorithm. 
	
	\subsection{Network evaluation and applications}
	
	Network evaluation is to evaluate the elements in the network. The degree and betweenness centralities are two popular network evaluation measures to evaluate the importance of vertices in the network. However, the degree centrality neglects the structural significance of a vertex while the betweenness centrality requires the global information of the network. Thus, both measures have defects in some applications. Liu et al. \cite{abii2016} proposed a PA based on the flow-conductivity model to evaluate the importance of vertices in the network. Their algorithm considers the degree of a focal node and the way neighbors are connected among themselves. Thus, their algorithm overcomes the shortcomings of degree and betweenness centralities. They have applied their algorithm to various empirical networks and showed its advantages over the popular measures.
	
	\subsection{The supply chain network design problem}
	
	A supply chain network is a network of manufacturers, distribution centers, and customers. The supply chain network design problem is to design supply chain networks to minimize the cost of transporting products from manufacturers to customers. Recently, Zhang et al. \cite{aips2016} proposed a PA based on the flow-conductivity model to challenge this problem. Their algorithm converges to new equilibrium states quickly when the network conditions are changed. Therefore, since the real-world supply chain networks are highly dynamic, their algorithm may have a better performance than traditional techniques. 
	
	\subsection{The transportation network design problem} 
	
	Transportation is a ubiquitous issue for human society as well as for biological systems. Therefore, the architecture, efficiency and other characteristics of the transportation networks have been attracting increasing scientific interest. Normally, transportation networks are required to have low cost, high efficiency, fault-tolerance, etc. Thus, transportation network design is a multi-objective optimization problem, which is hard to solve and has attracted intensive notices in recent years. Watanabe et al. \cite{toir2011} first used PAs to design railroad networks with balanced performances. Later, they designed transportation networks with fluctuating traffic distributions \cite{tnwf2014}. Besides their work, Houbraken \cite{ftnd2013} proposed a PA to design fault-tolerant transportation networks. Both their algorithms are based on the flow-conductivity model. There are some other types of PAs that have also been used to design transportation networks. For example, Jones and Becker et al. \cite{tpsm2011,aoba2013} proposed PAs based on the multi-agent model to do this. Moreover, besides the attempts to design transportation networks, Zhang et al. \cite{abaf2014} have also proposed a PA to identify critical parts in transportation networks.

	\section{Discussion} \label{Section: Discussion}

	PAs have the potential to solve various network optimization problems. There are three steps to develop them, which are conducting experiments to reveal Physarum intelligence, modeling Physarum intelligence, and customizing the model for applications. Many experiments have been done to reveal Physarum's intelligence for network optimization. These experiments show that the protoplasmic flow in the body of Physarum plays a great role in developing this intelligence. Thus, simulating the protoplasmic flow is essential in modeling Physarum intelligence. There are mainly four types of Physarum-inspired networking models, which are the flow-conductivity model, the cellular model, the multi-agent model, and the shuttle streaming model. Various PAs based on these models have been proposed to challenge network optimization problems.
	
	However, hardly any theoretical work has been done to support these challenges so far. For example, PAs based on the flow-conductivity model have challenged multiple versions of the Steiner tree problems, but rare work has been done to provide these PAs theoretical bases. As a consequence, it is hard to improve their performance. Hence, we suggest two types of future work, one is to provide PAs more theoretical bases, and the other one is to improve their performances. We have identified two fundamental questions: 1) What are the characteristics of Physarum networks? 2) Why can Physarum solve some network optimization problems? Answering these two questions is essential to the future development of Physarum-inspired network optimization. 
	
	Moreover, most existing PAs are still computationally too expensive for many network optimization problems. Therefore, minimizing their time complexities is another work worthy doing in the future. On the other hand, we observe that we can improve the performances of PAs not only by focusing on the Physarum-inspired techniques, but also by incorporating other bio-inspired techniques. For example, Sun et al. \cite{faib2016} incorporated the evolutionary computation technique to solve the node-weighted Steiner tree problem; Zhang et al. \cite{auos2014} incorporated the ant colony optimization algorithm to solve the traveling salesman problem; Liang et al. \cite{anga2016} incorporated the genetic algorithm to solve the multicast routing problem. All these hybrid PAs achieved a better performance than the original PAs. Thus, developing hybrid PAs is also recommended. With more solid theoretical bases and more available techniques, we believe that PAs can solve network optimization problems that are beyond the reach of traditional techniques in the future.

	
	\section{Conclusion}
	The existing researches on Physarum-inspired network optimization are still immature and far from being fully recognized. A major reason is that these researches have not been well organized so far. In this paper, we address this issue by summarizing and analyzing the existing Physarum-inspired networking models and the challenged network optimization problems and applications. We identify two fundamental questions: 1) What are the characteristics of Physarum networks? 2) Why can Physarum solve some network optimization problems? Answering these two questions is essential to the future development of Physarum-inspired network optimization.


	\bibliographystyle{ieeetr}
	\bibliography{YahuiBibIEEE}

\end{document}